\documentclass{webofc}
\usepackage[varg]{txfonts}   
\woctitle{epj}
\begin{document}
\title{Review of Recent Measurements of Meson and Hyperon Form Factors}

\author{Kamal K. Seth\inst{1}\fnsep\thanks{\email{kseth@northwestern.edu}} }

\institute{Northwestern University, Evanston, IL, 60208, USA}

\abstract{%
  A review of recent precision measurements of the electromagnetic form factors of the mesons, pion and kaon, and the hyperons, $\Lambda^0$, $\Sigma^0$, $\Sigma^+$, $\Xi^-$, $\Xi^0$, $\Omega^-$, at large timelike momentum transfers is presented.  Evidence is found for diquark correlations in $\Lambda^0$, $\Sigma^0$ hyperons.
}
\maketitle

It is generally agreed that electromagnetic form factors at large momentum transfer provide some of the best insight into the structure of a hadron. 
Four-momentum transfers defined as 
\begin{equation}
Q^2(\text{4~mom}) = q^2(\text{3~mom})_\text{space} - (\text{energy})_\text{time}
\end{equation}
can be \textit{positive and spacelike}, or \textit{negative and timelike}.  I am going to talk about form factors for timelike momentum transfers as measured via the reactions $e^+e^- \to \text{hadron-antihadron}$.

In 1960, the first proposals for electron-positron colliders were being considered at SLAC and Frascati. In anticipation of these, Cabibbo and Gatto wrote two classic papers~\cite{cabibbo} pointing out that these colliders would provide the unique opportunity to measure timelike form factors of any hadrons, mesons and baryons. Only 50 years later, we are now realizing the full promise of the vision of Cabibbo and Gatto  in the measurements I am reporting here.

Almost no experimental data with any precision existed before 2000 for pion and kaon spacelike or timelike form factors for $|Q^2| > 5$~GeV$^2$.
Recently, we made the first measurements of the form factors of pions and kaons with high precision for the large momentum transfers of $|Q^2|=14.2$ and 17.4~GeV$^2$.  Since these have been published~\cite{cleo-ff}, although I discussed these in detail in my talk, the space limit allows me to only present the results here.

\begin{figure}
\centering
\includegraphics[width=2.6in,clip]{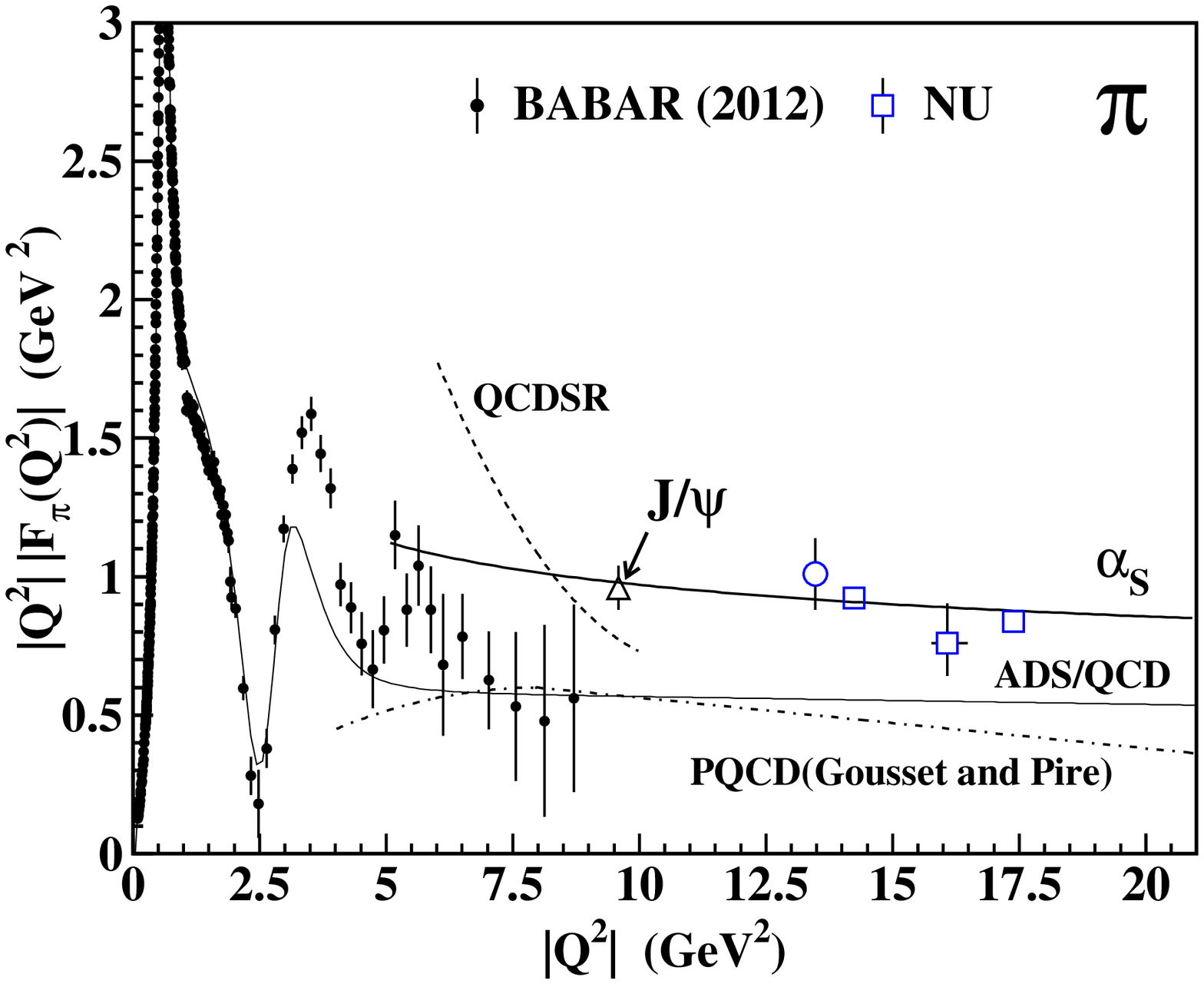}
\includegraphics[width=2.6in,clip]{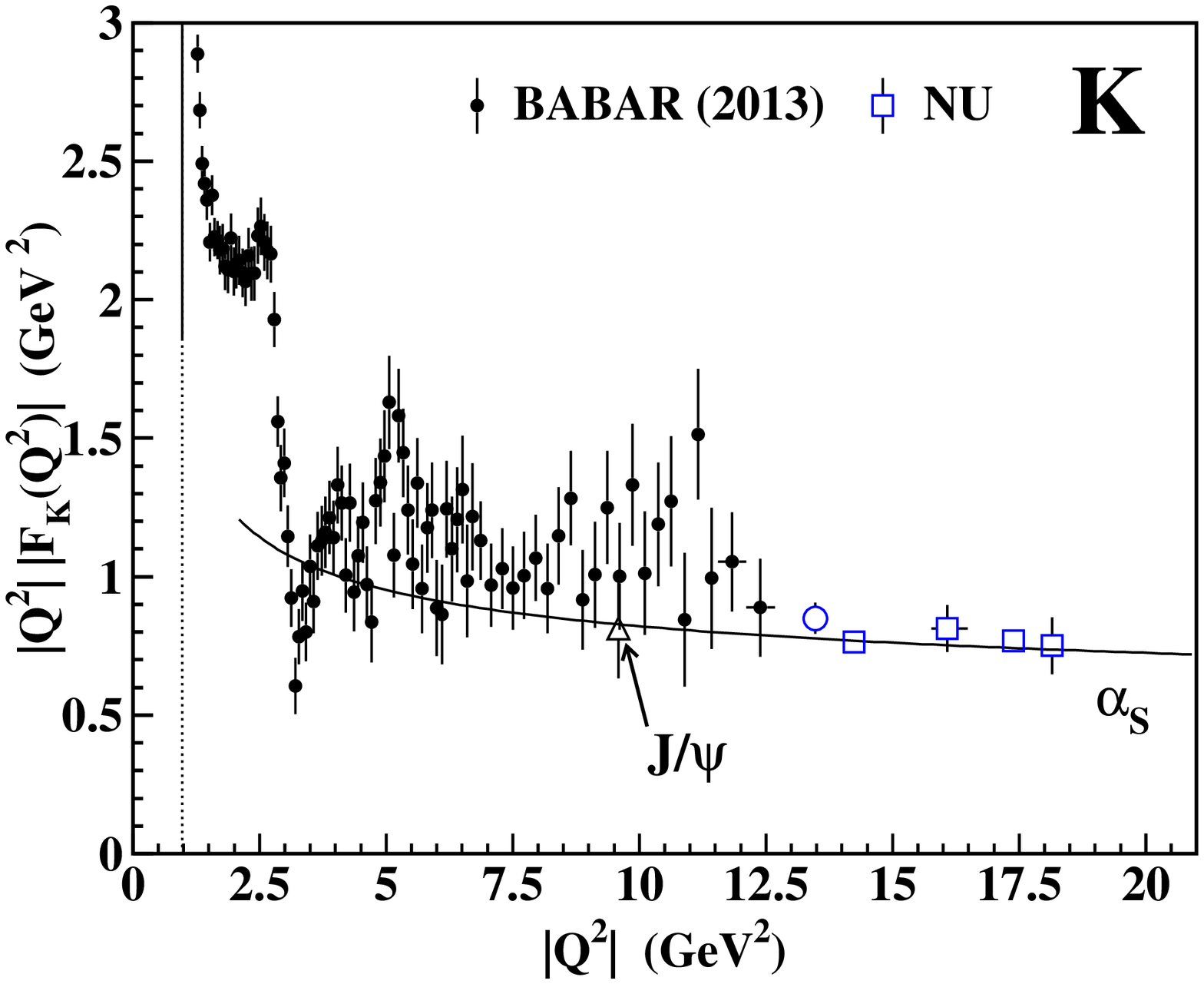}
\caption{Results for pion and kaon timelike form factors.}
\label{fig:pik}       
\end{figure}

The important experimental results are presented in Table~\ref{tbl:pik} and displayed in Fig.~\ref{fig:pik}.
\begin{enumerate}
\item There is a remarkable agreement of the form factors for both pions and kaons with the dimensional counting rule prediction of QCD, that $|Q^2|F_{\pi,K}$ are nearly constant, varying with $|Q^2|$ only weakly as $\alpha_S(|Q^2|)$.
\item The existing theoretical predictions for pions underpredict the magnitude of  $F_\pi(|Q^2|)$ at large $|Q^2|$ by large factors, $\ge2$.
\item The big surprise is that while pQCD predicts that $F_\pi/F_K=(f_\pi/f_K)2=0.67\pm0.01$, we find: 
\begin{equation}
F_\pi / F_K = 1.21 \pm 0.03,~\text{at}~|Q^2| = 14.2 \text{GeV}^2, \quad
F_\pi / F_K = 1.09 \pm 0.04,~\text{at}~|Q^2| = 17.4 \text{GeV}^2.
\end{equation}
\end{enumerate}
It has been suggested that this dramatic disagreement may be due to the kaon wave function being different from that of the pion due to $SU(3)$ breaking, the strange quark in the kaon having a 27 times larger mass than the up/down quarks in the pion.

\begin{table}
\centering
\caption{Results for pion and kaon timelike form factors.}
\label{tbl:pik}

\begin{tabular}{ccc}
\hline
 $Q^2$ (GeV$^2$) & $Q^2F_\pi(|Q^2|)$ (GeV$^2$) & $Q^2F_K(|Q^2|)$ (GeV$^2$) \\
\hline
 14.2 & $0.92\pm0.04$ & $0.76\pm0.02$ \\
 17.4 & $0.84\pm0.03$ & $0.77\pm0.03$ \\
\hline
\end{tabular}

\end{table}

Quarks were not even invented when Cabibbo and Gatto suggested that measurement of hyperon form factors would be interesting.  In the present context of QCD and the quark-gluon structure of hadrons, it is particularly interesting to measure form factors of hyperons which may be expected to reveal the effects of $SU(3)$ breaking, as successively one, two, and three of the up/down quarks in the nucleon are replaced by strange quarks in ($\Lambda,\Sigma$), $\Xi$, and $\Omega$, respectively.  The interest is further enhanced at large momentum transfers at which deeper insight is obtained into possible short-range correlations between the quarks.  As Wilczek~\cite{wilczek1} has pointed out, ``several of the most profound aspects of low-energy QCD dynamics are connected to diquark correlations,'' and the differences in quark-quark configurations between different hyperons make them an ideal laboratory to study diquark correlations.

The $e^+e^-\to \text{hyperon-antihyperon}$ cross sections were expected to be very small, and no experimental measurements were reported for 47~years after Cabibbo and Gatto's papers.  In 2007, BaBar reported~\cite{babar-ff} form factor measurements for $\Lambda$ and $\Sigma^0$ using the ISR method, but with statistically significant results limited to $|Q^2|\lesssim6$~GeV$^2$.
And now we have measured form factors of all hyperons, $\Lambda^0$, $\Sigma^0$, $\Sigma^+$, $\Xi^-$, $\Xi^0$, $\Omega^-$ for the first time with good precision at the large momentum transfer of $|Q^2|=14.2$~GeV$^2$.

The only existing theoretical study of hyperon form factors at large momentum transfers is due to K\"orner and Kuroda~\cite{kornerkuroda}, who in 1977 made predictions of form factor cross sections for nucleons and hyperons for timelike momentum transfers ranging from threshold to $|Q^2|=16$~GeV$^2$ in the framework of the Generalized Vector Dominance Model (GVDM). These predictions were not constrained by any experimental measurements, and as we shall see, they turn out to be factors 10 to 80 larger than what we measure.

 We use data taken with the CLEO-c detector at $\psi(3770)$, $\sqrt{s}=3.77$~GeV, with the integrated luminosity $\mathcal{L}=805$~pb$^{-1}$. 
Data taken at $\psi(3770)$ can only be used to determine hyperon form factors if it can be shown that the strong interaction yield of the hyperons at the resonance is very small.  We estimate it by using the pQCD prediction that the ratios of the branching fractions for the decay of any two vector resonances of charmonium to leptons (via a photon) and hadrons (via gluons) are identical, because both are proportional to the wave functions at the origin.  This relation was successfully used by us recently to measure form factors of pions and kaons at $\psi(3770)$ and $\psi(4160)$~\cite{cleo-ff}.  In the present case, it leads to 
\begin{equation}
\frac{\mathcal{B}(\psi(3770)\to\text{gluons}\to\text{hyperons})}{\mathcal{B}(J/\psi,\psi(2S)\to\text{gluons}\to\text{hyperons})}  = \frac{\mathcal{B}(\psi(3770)\to\text{photon}\to\text{electrons})}{\mathcal{B}(J/\psi,\psi(2S)\to\text{photon}\to\text{electrons})}
\end{equation}
Using the measured branching fractions for $J/\psi$~\cite{pdg} and $\psi(2S)$~[9,present] we find that $\mathcal{B}(\psi(3770)\to\text{hyperons})<4\times10^{-7}$ for all hyperons, and they lead to the expected number of events, $1.3~p$, $0.9~\Lambda^0$, $0.2~\Sigma^+,\Sigma^0,~\Xi^-$, $0.05~\Xi^0$, and $0.03~\Omega^-$ for resonance decays of $\psi(3770)$ in the present measurements.  In other words, the contributions of strong decays are negligibly small in all decays, and the observed events arise from form factor decays.

We also use CLEO-c data taken at $\psi(2S)$, $\sqrt{s}=3.686$~GeV, with luminosity $\mathcal{L}=48$~pb$^{-1}$, which corresponds to $N(\psi(2S))=24.5\times10^6$, to measure the branching fractions for the decays $\psi(2S)\to B\overline{B}$.  The large yield from resonance production of $B\overline{B}$ pairs from $\psi(2S)$ enables us to test the quality of our event selection criteria, and to determine contributions to systematic uncertainties.

For both $\psi(2S)$ and $\psi(3770)$ decays we reconstruct the hyperons in their following major decay modes (with branching fractions \cite{pdg} listed in parentheses): $\Lambda^0\to p\pi^-$ (63.9\%), $\Sigma^+\to p\pi^0$ (51.6\%), $\Sigma^0 \to \Lambda^0 \gamma$ (100\%), $\Xi^-\to\Lambda^0\pi^-$ (99.9\%), $\Xi^0\to\Lambda^0\pi^0$ (99.5\%), $\Omega^- \to \Lambda^0 K^-$ (67.8\%).  We find that reconstructing back-to-back hyperons and anti-hyperons whose decay vertices are separated from the interaction point results in essentially background free spectra.

I will not go into the nitty-gritty of particle identification here.  Suffice it to say that using energy loss ($dE/dx$) in the CLEO drift chamber, and the log-likelihood, $L^\text{RICH}$, information from the RICH detector, we first identify single hyperons, and then construct hyperon-antihyperon pairs.  Both steps are illustrated in Fig.~2 for $\psi(2S)$ resonance decays which have large yields.

To determine the reconstruction efficiency of the above event selections, we generate Monte Carlo events using a GEANT-based detector simulation.  For the decay of $\psi(2S)$ to spin--1/2 baryon pairs ($\Lambda,\Sigma,\Xi$), we generate events with the expected angular distribution of $1+\cos^2 \theta$. 
For the spin--3/2 $\Omega^-$ hyperon, we generate events with the angular distribution  $[\sin\frac{\theta}{2}(1+3\cos\theta) + \cos\frac{\theta}{2}(1-3\cos\theta)]^2$ expected for spin~$1\to3/2+3/2$.

\begin{figure}[!tb]
\begin{center}
\includegraphics[width=2.7in]{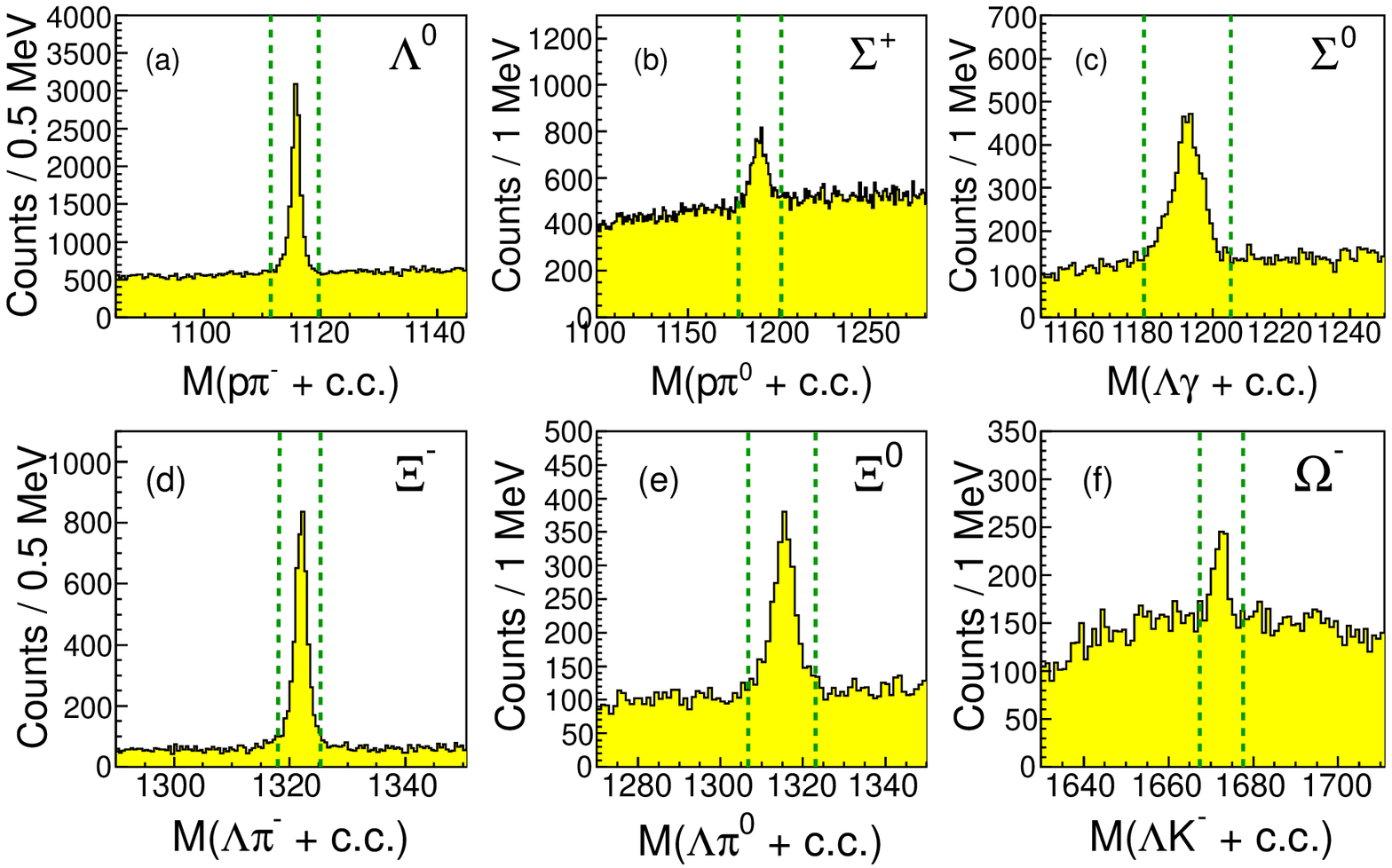}
\includegraphics[width=2.7in]{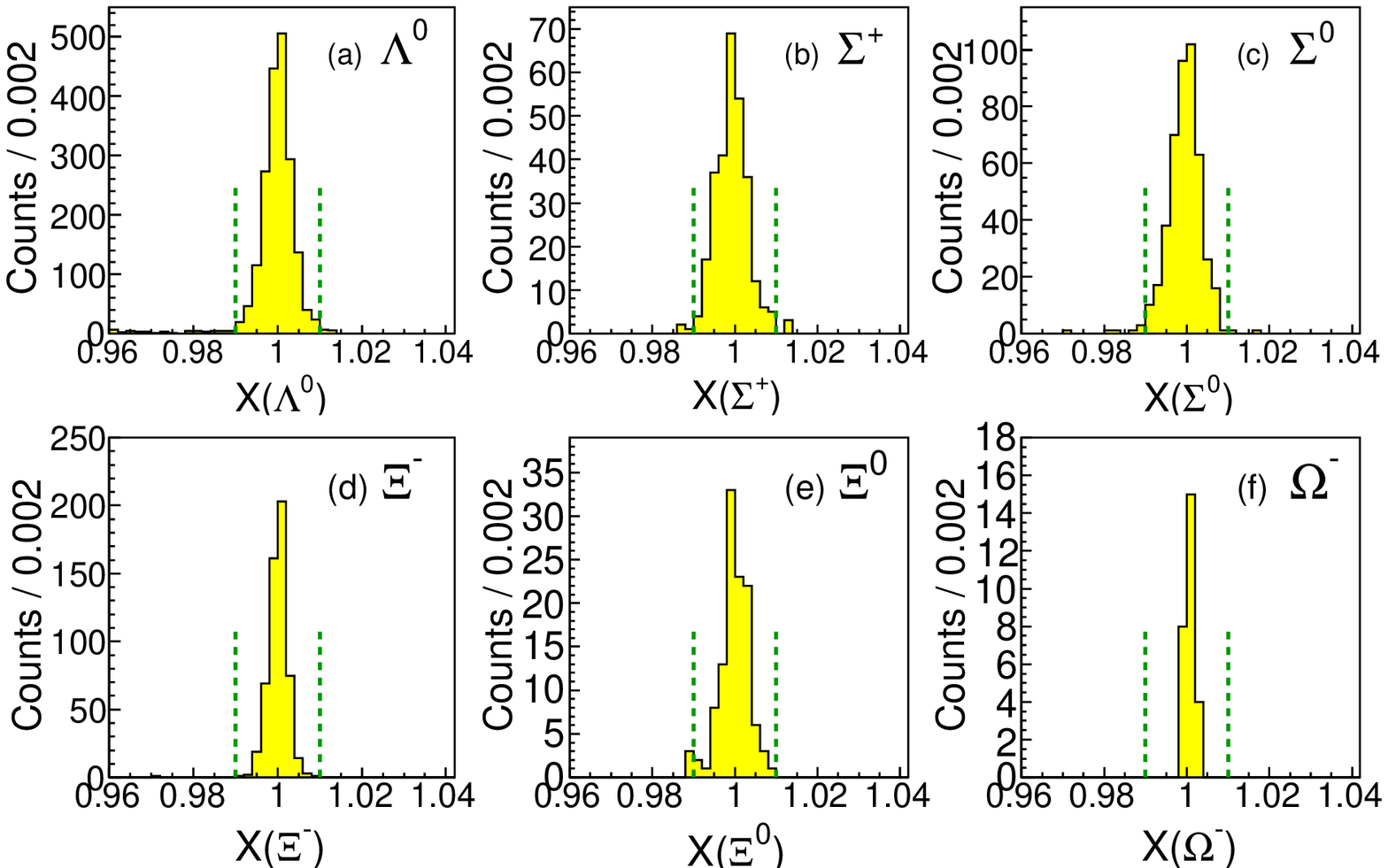}
\end{center}

\caption{(left) Invariant masses of reconstructed final states in $\psi(2S)$ data.  Single hyperons are accepted in the regions defined by vertical dashed lines.  (right) Distributions of baryon--antibaryon events as function of, $X(B)\equiv(E(B)+E(\overline{B}))/\sqrt{s}$, in $\psi(2S)$ data.  The vertical lines indicate the signal region $X=0.99-1.01$.}

\label{fig:psi2smass}
\end{figure}

The single hyperon mass spectra for $\psi(2S)$ decays are shown in Fig.~\ref{fig:psi2smass}(left), for hyperons with $E(B)/E_\text{beam}>0.95$.
Clear peaks are seen for the reconstructed hyperons with varied levels of background and peak widths depending on the final state particles.

\begin{table}[!tb]

\centering
\caption{Cross section and branching fraction results for $\psi(2S)\to B\overline{B}$.}
\label{tbl:psi2sbr}

\begin{tabular}{l|cccc}
\hline \hline
$\mathcal{B}$ & $N_\text{data}$ & $\epsilon_B$ (\%) & $\sigma_B$ (pb) & $\mathcal{B}\times10^4$  \\
\hline
$p$         & $4475(78)$  & 63.1  & $196(3)(12)$ & $3.08(5)(18)$  \\
$\Lambda^0$ & $1901(44)$  & 20.7  & $247(6)(15)$  & $3.75(9)(23)$  \\
$\Sigma^0$  & $ 439(21)$  & 7.96  & $148(7)(11)$  & $2.25(11)(16)$ \\
$\Sigma^+$  & $ 281(17)$  & 4.54  & $165(10)(11)$ & $2.51(15)(16)$ \\
$\Xi^-$     & $ 548(23)$  & 8.37  & $176(8)(13)$  & $2.66(12)(20)$ \\
$\Xi^0$     & $ 112(11)$  & 2.26  & $135(13)(10)$ & $2.02(19)(15)$ \\
$\Omega^-$  & $  27(5)$   & 2.32  & $31(6)(3)$    & $0.47(9)(5)$   \\

\hline \hline
\end{tabular}

\end{table}

The second step consists of constructing  baryon--antibaryon pairs. The distributions of these $B\overline{B}$ pairs is shown in Fig.~2(right) for $\psi(2S)$ decays as a function of $X(B)\equiv[E(B)+E(\overline{B})]/\sqrt{s}$, which should peak at $X(B)=1$. Clear peaks are seen for all decays with essentially no background.  We define the signal region as $X(B)=0.99-1.01$, with numbers of events in it as $N_\text{data}$.  The estimated number of events due to form factor contribution under the peaks is found to be negligable, being less than 1\% in all cases.   We calculate the radiative correction, $(1+\delta)$, using the method of Bonneau and Martin~\cite{bonneaumartin}. We obtain $(1+\delta)=0.77$ within 1\% for all baryons for both $\psi(2S)$ and $\psi(3770)$.  The Born cross sections are calculated as $\sigma_B=N_\text{data}/\epsilon_B\,\mathcal{L}(\psi(2S))\,(1+\delta)$, and the branching fractions as $\mathcal{B}(\psi(2S)\to B\overline{B}) = N_\text{data}/\epsilon_B N(\psi(2S))$.  The results are summarized in Table~\ref{tbl:psi2sbr}, including those for $\psi(2S)\to p\bar{p}$.  The first uncertainties in $\sigma_B$ and $\mathcal{B}$ are statistical, and the second uncertainties are estimates of systematic uncertainties. 
Our results for $\psi(2S)$ branching fractions are in agreement with the PDG averages~\cite{pdg} and previous small luminosity CLEO results~\cite{cleo-psi2sbb}, and have generally smaller errors.

\begin{figure}[!tb]
\begin{center}
\includegraphics[width=2.7in]{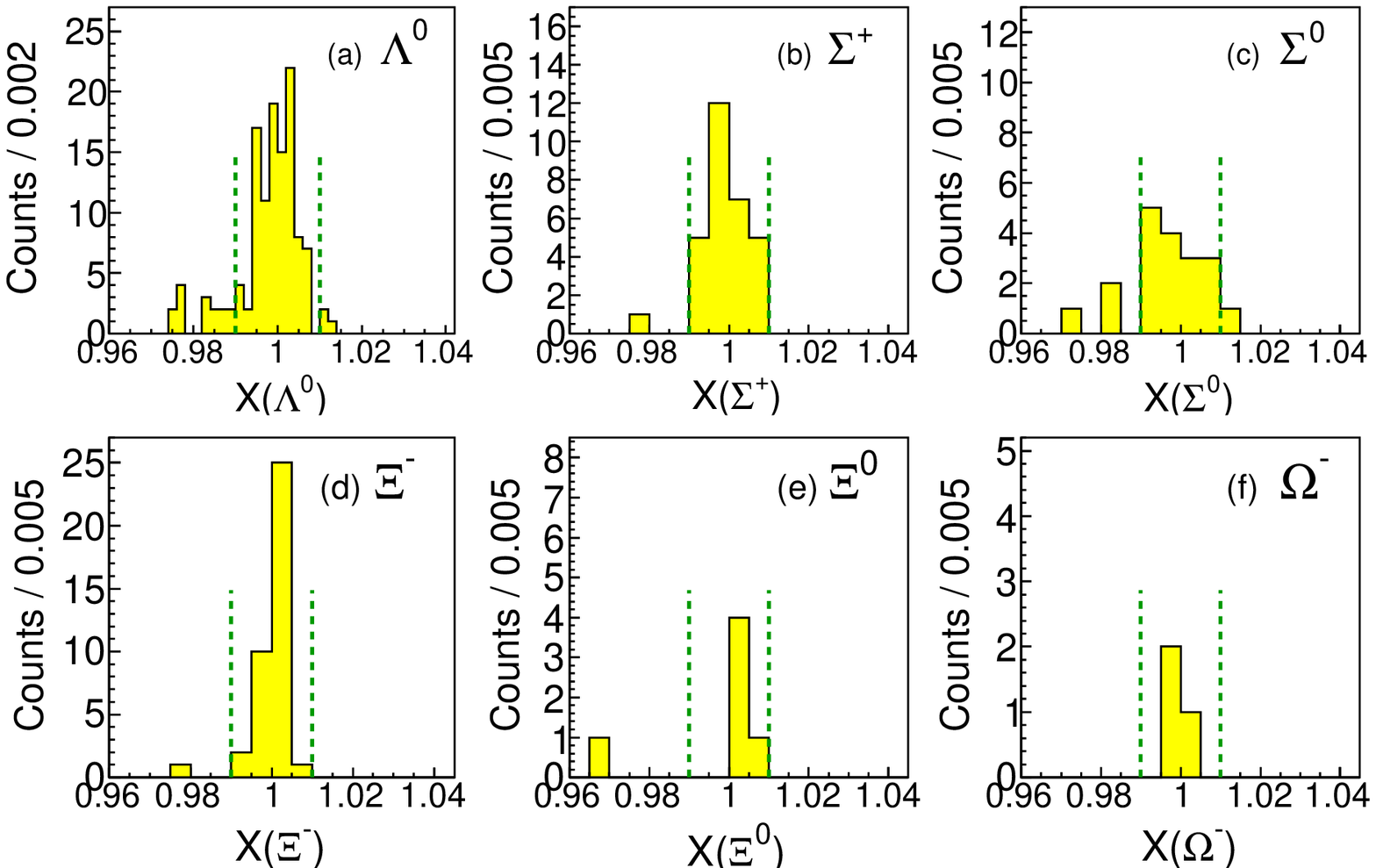}
\raisebox{10pt}{\includegraphics[width=2.3in]{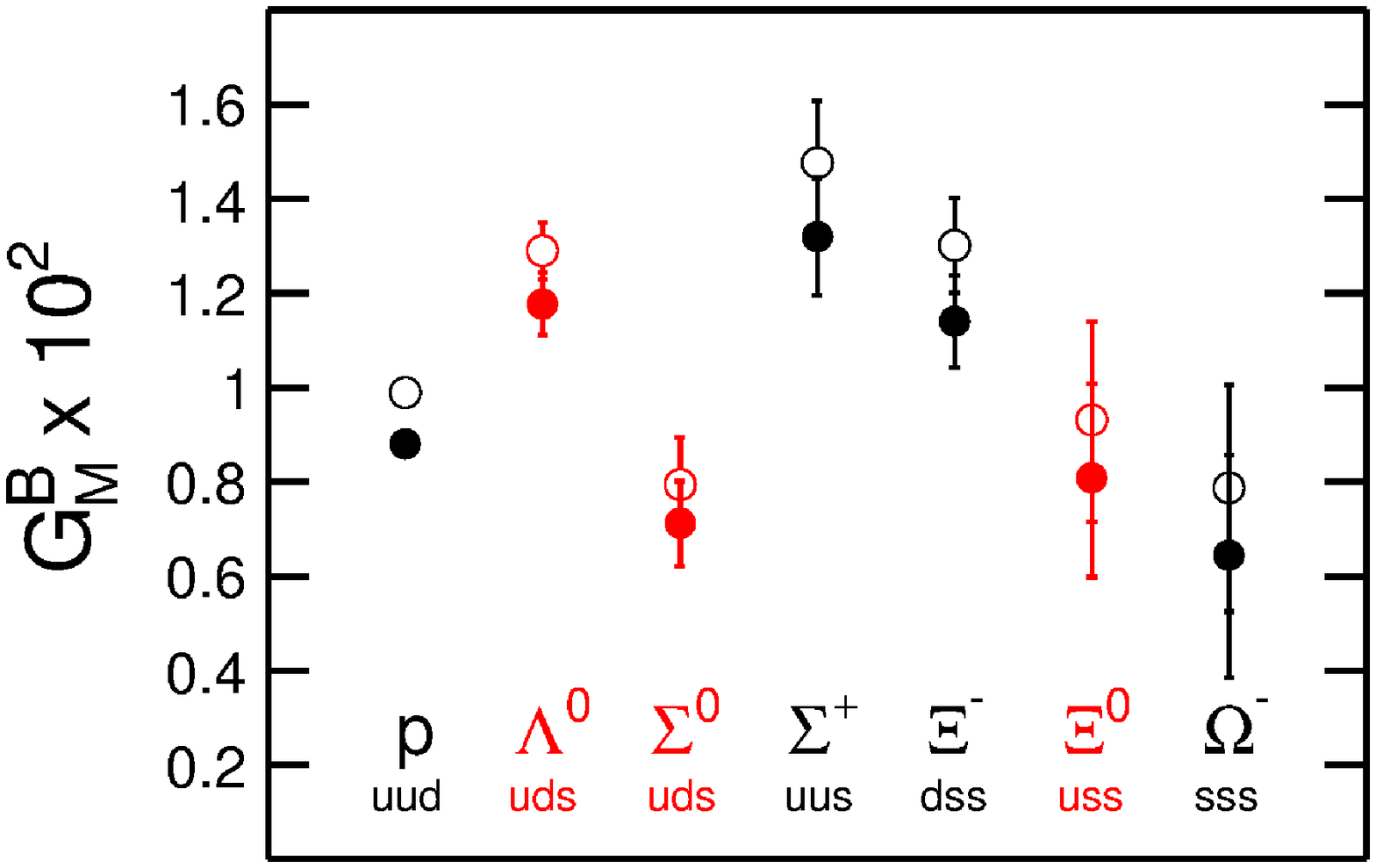}}
\end{center}

\caption{(left) Distributions of baryon-antibaryon scaled energy, $X(B)\equiv(E(B)+E(\overline{B}))/\sqrt{s}$, in $\sqrt{s}=3770$~MeV data.  The vertical lines indicate the signal region $X=0.99-1.01$.  (right) Magnetic form factors $|G_M^B|\times10^2$ for proton and hyperons.  The closed circles correspond to the assumption $G_M^B=G_E^B$, and the open circles to the assumption $G_E^B=0$.}
\label{fig:psi3770xb}
\end{figure}

We apply the same event selections to the $\psi(3770)$ decays as we do for $\psi(2S)$ decays.  The $X(B)$ distributions for $\psi(3770)$ form factors decays are shown in Fig.~\ref{fig:psi3770xb}(left).  Clear peaks are seen for each decay mode with yields ranging from 105 for $\Lambda^0\overline{\Lambda^0}$ to 3 for $\Omega^-\overline{\Omega^-}$.  The few events seen in the neighborhood of $X\approx0.98$ are consistent with being from the decay of $\psi(2S)$ populated by initial state radiation (ISR). The number of events, $N_\text{ff}$, in the region $X(B)=0.99-1.01$, are used to calculate the cross sections as, $\sigma_0(e^+e^- \to B\overline{B}) = N_\text{ff}/(1+\delta)\epsilon_B\,\mathcal{L}(3770)$, where $\epsilon_B$ are the MC-determined efficiencies at $\sqrt{s}=3770$~MeV, $(1+\delta)=0.77$ is the radiative correction, and $\mathcal{L}(3770)=805$~pb$^{-1}$ is the luminosity at $\sqrt{s}=3770$~MeV.

For the spin--1/2 baryons, the proton and the hyperons $\Lambda$, $\Sigma$, and $\Xi$, the well known relation between the cross sections and the magnetic from factor $G_M^B(s)$, and the electric form factor $G_E^B(s)$ is
\begin{equation}
\sigma_0^B = \left( \frac{4\pi\alpha^2\beta_B}{3s}\right) \left[ |G_M^B(s)|^2 + (2m_B^2/s)|G_E^B(s)|^2 \right]
\end{equation}
where $\alpha$ is the fine structure constant, $\beta_B$ is the velocity of the baryon in the center-of-mass system, and $m_B$ is its mass.  The statistics of the present measurements do not allow us to determine $|G_M^B|$ and $|G_E^B|$ separately.  We therefore evaluate $|G_M^B(s)|$ under two commonly used extreme assumptions,  $|G_M^B(s)|/|G_E^B(s)|=0$, and 1.  The results for $|G_E|=|G_M|$ are shown in Table~\ref{tbl:hypff}.  
The efficiencies for the $G_M$ and $G_E$ components are determined assuming $1+\cos^2\theta$ and $\sin^2\theta$ angular distributions, respectively.
As shown in Fig.~3~(right), the values of $G_M^B(s)$ derived with the assumption $G_E^B=0$ are found to all be nearly $\sim12(2)\%$ larger than those for $G_E^B(s)=G_M^B(s)$.

For the spin--3/2 $\Omega^-$, there are four form factors, $G_{E0}$, $G_{E2}$, $G_{M1}$, and $G_{M3}$.  Following K\"orner and Kuroda~\cite{kornerkuroda}, Eq.~2 is valid if it is understood that $G_M^B$ includes the contributions of both magnetic quadrupole ($G_{M1}$) and octopole ($G_{M3}$) form factors, and $G_E^B$ includes the contributions of both electric dipole ($G_{E0}$) and quadrupole ($G_{E2}$) form factors.

We evaluate systematic uncertainties due to various sources for each hyperon pair, and add the contributions from the different sources together in quadrature.  The systematic uncertainties total to 6.1\% for $\Lambda^0$, 7.3\% for $\Sigma^0$, 6,4\% for $\Sigma^+$, 7.5\% for $\Xi^-$, 7.3\% for $\Xi^0$, and 10.2\% for $\Omega^-$.

\begin{table}[!tb]

\centering 
\caption{Results for proton and hyperon form factors at $|Q^2|=14.2$~GeV$^2$, assuming $|G_E|=|G_M|$.} 
\label{tbl:hypff}

\begin{tabular}{l|ccccc}
\hline \hline

$B$ & \multicolumn{1}{c}{$N_\text{ff}$} & $\epsilon_B$, \% & $\sigma_0^B$, pb & $|G_M^B|\!\times\!10^2$  & $|G_M^B/\mu_B|\!\times\!10^2$   \\
\hline
$p$         & $215(15)$ & $71.3$ & $0.46(3)(3)$ & $0.88(3)(2)$ & $0.31(1)(1)$ \\
$\Lambda^0$ & $105(10)$  & $21.1$ & $0.80(8)(5)$  & $1.18(6)(4)$  & $1.93(9)(6)$ \\
$\Sigma^0$  &  $ 15(4)$  & $8.36$ & $0.29(7)(2)$  & $0.71(9)(3)$  & $0.91(11)(3)$ \\
$\Sigma^+$  & $ 29(5) $  & $4.68$ & $0.99(18)(6)$ & $1.32(13)(4)$ & $0.54(5)(2)$ \\
$\Xi^-$     & $ 38(6) $  & $8.69$ & $0.71(11)(5)$ & $1.14(9)(4)$  & $1.75(14)(7)$ \\
$\Xi^0$     &  $  5^{+2.8}_{-2.3}$ & $2.30$ & $0.35^{+0.20}_{-0.16}(3)$ & $0.81(21)(3)$ & $0.65(17)(2)$ \\
$\Omega^-$  &  $  3^{+2.3}_{-1.9}$ & $2.94$ & $0.16^{+0.13}_{-0.10}(2)$ & $0.64^{+0.21}_{-0.25}(3)$ & $0.32^{+0.11}_{-0.13}(2)$ \\
\hline \hline
\end{tabular}
\end{table}

Since no modern theoretical predictions for timelike form factors of hyperons at large momentum transfers exist, we can only discuss our experimental results qualitatively.  Following are the main observations:
\begin{enumerate}
\item[(a)] The form factor cross sections in Table~\ref{tbl:hypff} are 150 to 500 times smaller than the resonance cross sections in Table~\ref{tbl:psi2sbr}.   Clearly, larger statistics measurements of the form factors would be highly desirable.

\item[(b)] As illustrated in Fig.~3(right), the measured values of $|G_M^B|$ vary rather smoothly by approximately a factor two, except for $G_M(\Sigma^0)$.  

\item[(c)] It is common practice to quote spacelike form factors for protons as $G^p_M(s)/\mu_p$, based on normalization at $|Q^2|=0$.  We note that there is no evidence for the proportionality of $G_M^B(s)$ to $\mu_B$ for hyperons.  As listed in Table~\ref{tbl:hypff}, $G_M^B(s)/\mu_B$ vary by more than a factor~4 for the different baryons.
\end{enumerate}
The most significant result of the present measurements is that $G_M(\Lambda^0)$ is a factor $1.66(24)$ larger than $G_M(\Sigma^0)$, although $\Lambda^0$ and $\Sigma^0$ have the same $uds$ quark content. We note that $\Sigma^0$ and $\Lambda^0$ differ in their isospin, with $I(\Sigma^0)=1$, and $I(\Lambda^0)=0$.  Since only up and down quarks carry isospin, this implies that the pair of up/down quarks in $\Lambda^0$ and $\Sigma^0$ have different isospin configurations.  This forces different spin configurations in $\Lambda^0$ and $\Sigma^0$.  In $\Lambda^0$ the $ud$ quarks have antiparallel spins coupled to $S=0$, whereas in $\Sigma^0$ they couple to $S=1$.  The spatial overlap in the $S=0$ configuration in $\Lambda^0$ is stronger than in the $S=1$ configuration in $\Sigma^0$, and our measurement at large $|Q^2|$ is particularly sensitive to it.

Recently, Wilczek and colleagues~\cite{wilczek1,jaffewilczek,wilczek2} have emphasized the importance of diquark correlations in low-energy QCD dynamics, and have pointed out that for the non-strange quarks the favorable diquark configuration with attraction is the spin-isospin singlet, making what Wilczek calls a ``good'' diquark in $\Lambda^0$ as opposed to the repulsive spin-isospin triplet configuration in $\Sigma^0$.  This results in a significantly larger cross section for the formation of $\Lambda^0$ than $\Sigma^0$, as anticipated by Selem and Wilczek~\cite{wilczek2}.  We measure $\sigma(\Lambda^0)/\sigma(\Sigma^0)\approx3$, and this results in the factor 1.66 larger form factor for $\Lambda^0$ than $\Sigma^0$. We believe that our observation of the large difference in the form factors of $\Lambda^0$ and $\Sigma^0$ is indeed due to significant ``good'' diquark correlation in $\Lambda^0$, and it constitutes an important example of significant diquark correlations in baryons.

%
%
%

\end{document}